\pdfoutput=1  

\def\virg#1{\textquoteleft #1\textquoteright}

\def\tn{\textnormal{-}}

\def\aco{\aca{\qo}}

\def\ho{\rb{\qo}}

\def\li{$\mcr{\imath}$}

\def\bo{$\mathit{\breve{o}}$}

\def\bds#1{\mathbf{#1}}
\def\acap{\\ \nonumber \\}

\def\rfr#1{Equation\,(\ref{#1})}
\def\rfrs#1#2{Equations\,(\ref{#1})--(\ref{#2})}

\def\dert#1#2{\frac{{{\mathit{d}}}{#1}}{{{\mathit{d}}}{#2}}} 
 
\def\eqi{\begin{equation}}
\def\eqf{\end{equation}}
\def\eqia{\begin{eqnarray}}
\def\eqfa{\end{eqnarray}}
\def\rp#1#2{\frac{#1}{#2}}
\def\lb#1{\label{#1}}

\def\bds#1{\boldsymbol{#1}}
\def\nk{n_\mathrm{K}}

\def\ton#1{\left(#1\right)}
\def\qua#1{\left[#1\right]}
\def\grf#1{\left\{#1\right\}}
\def\ang#1{\left\langle #1\right\rangle}

\documentclass[a4paper, 11pt]{article}

\usepackage[english]{babel}
\usepackage[LGR,T1]{fontenc}
\usepackage[utf8]{inputenc}
\usepackage{w-greek}
\usepackage[nointegrals]{wasysym}
\usepackage{authblk}
\usepackage{abstract}
\usepackage{amsmath}
\usepackage{amsthm}
\usepackage{amscd}
\usepackage{amssymb}
\usepackage{dsfont}
\usepackage{xcolor}
\usepackage{graphicx}
\usepackage{epsfig}
\usepackage{xr-hyper}
\usepackage[pdfstartview=XYZ,
bookmarks=true,
colorlinks=true,
linkcolor=blue,
urlcolor=blue,
citecolor=blue,
pdftex,
bookmarks=true,
linktocpage=true, 
hyperindex=true
]{hyperref}
\usepackage{starfont}
\usepackage{txfonts}
\usepackage[mathlines]{lineno}
\usepackage{tabularx}
\usepackage{booktabs}
\usepackage{fontawesome5}

\usepackage[round]{natbib} 
\allowdisplaybreaks[1]

\makeatletter
\DeclareRobustCommand\ref{%
    \@ifstar\@refstar\T@ref
  }%
  \DeclareRobustCommand\pageref{%
    \@ifstar\@pagerefstar\T@pageref
  }%
 \makeatother

\DeclareSymbolFont{greekletters}{LGR}{\familydefault}{m}{n}
\DeclareMathSymbol{\qA}{\mathord}{greekletters}{65}
\DeclareMathSymbol{\qB}{\mathord}{greekletters}{66}
\DeclareMathSymbol{\qG}{\mathord}{greekletters}{71}
\DeclareMathSymbol{\qD}{\mathord}{greekletters}{68}
\DeclareMathSymbol{\qE}{\mathord}{greekletters}{69}
\DeclareMathSymbol{\qZ}{\mathord}{greekletters}{90}
\DeclareMathSymbol{\qEt}{\mathord}{greekletters}{72}
\DeclareMathSymbol{\qTh}{\mathord}{greekletters}{74}
\DeclareMathSymbol{\qI}{\mathord}{greekletters}{73}
\DeclareMathSymbol{\qK}{\mathord}{greekletters}{75}
\DeclareMathSymbol{\qL}{\mathord}{greekletters}{76}
\DeclareMathSymbol{\qM}{\mathord}{greekletters}{77}
\DeclareMathSymbol{\qN}{\mathord}{greekletters}{78}
\DeclareMathSymbol{\qX}{\mathord}{greekletters}{88}
\DeclareMathSymbol{\qO}{\mathord}{greekletters}{79}
\DeclareMathSymbol{\qP}{\mathord}{greekletters}{80}
\DeclareMathSymbol{\qR}{\mathord}{greekletters}{82}
\DeclareMathSymbol{\qS}{\mathord}{greekletters}{83}
\DeclareMathSymbol{\qT}{\mathord}{greekletters}{84}
\DeclareMathSymbol{\qU}{\mathord}{greekletters}{85}
\DeclareMathSymbol{\qPh}{\mathord}{greekletters}{70}
\DeclareMathSymbol{\qCh}{\mathord}{greekletters}{81}
\DeclareMathSymbol{\qPs}{\mathord}{greekletters}{89}
\DeclareMathSymbol{\qOm}{\mathord}{greekletters}{87}
\DeclareMathSymbol{\qa}{\mathord}{greekletters}{97}
\DeclareMathSymbol{\qb}{\mathord}{greekletters}{98}
\DeclareMathSymbol{\qg}{\mathord}{greekletters}{103}
\DeclareMathSymbol{\qd}{\mathord}{greekletters}{100}
\DeclareMathSymbol{\qe}{\mathord}{greekletters}{101}
\DeclareMathSymbol{\qz}{\mathord}{greekletters}{122}
\DeclareMathSymbol{\qet}{\mathord}{greekletters}{104}
\DeclareMathSymbol{\qth}{\mathord}{greekletters}{106}
\DeclareMathSymbol{\qi}{\mathord}{greekletters}{105}
\DeclareMathSymbol{\qk}{\mathord}{greekletters}{107}
\DeclareMathSymbol{\ql}{\mathord}{greekletters}{108}
\DeclareMathSymbol{\qm}{\mathord}{greekletters}{109}
\DeclareMathSymbol{\qn}{\mathord}{greekletters}{110}
\DeclareMathSymbol{\qx}{\mathord}{greekletters}{120}
\DeclareMathSymbol{\qo}{\mathord}{greekletters}{111}
\DeclareMathSymbol{\qp}{\mathord}{greekletters}{112}
\DeclareMathSymbol{\qr}{\mathord}{greekletters}{114}
\DeclareMathSymbol{\fs}{\mathord}{greekletters}{99}       
\DeclareMathSymbol{\qs}{\mathord}{greekletters}{115}       
\DeclareMathSymbol{\qt}{\mathord}{greekletters}{116}
\DeclareMathSymbol{\qu}{\mathord}{greekletters}{117}
\DeclareMathSymbol{\qvf}{\mathord}{greekletters}{102}
\DeclareMathSymbol{\qch}{\mathord}{greekletters}{113}
\DeclareMathSymbol{\qps}{\mathord}{greekletters}{121}
\DeclareMathSymbol{\qom}{\mathord}{greekletters}{119}
\DeclareMathSymbol{\sui}{\mathord}{greekletters}{124}
\DeclareMathSymbol{\dig}{\mathord}{greekletters}{147}
\DeclareMathAccent{\uml}{\mathord}{greekletters}{34} 
\DeclareMathAccent{\umld}{\mathord}{greekletters}{35} 
\DeclareMathAccent{\mcr}{\mathord}{greekletters}{31}
\DeclareMathAccent{\aca}{\mathord}{greekletters}{39} 
\DeclareMathAccent{\ga}{\mathord}{greekletters}{96} 
\DeclareMathAccent{\rb}{\mathord}{greekletters}{60} 
\DeclareMathAccent{\smb}{\mathord}{greekletters}{62} 
\DeclareMathAccent{\ca}{\mathord}{greekletters}{126} 
\DeclareMathAccent{\carb}{\mathord}{greekletters}{64} 
\DeclareMathAccent{\casb}{\mathord}{greekletters}{92} 
\DeclareMathAccent{\aarb}{\mathord}{greekletters}{86} 
\DeclareMathAccent{\aasb}{\mathord}{greekletters}{94} 
\DeclareMathAccent{\garb}{\mathord}{greekletters}{67} 
\DeclareMathAccent{\gasb}{\mathord}{greekletters}{95} 

\DeclareTextSymbol{\textquoteleft}{LGR}{28}      
\DeclareTextSymbol{\textquoteright}{LGR}{29}     

\definecolor{orcidlogocol}{rgb}{0.65, 0.807, 0.223}

\newcommand{\orcid}[1]{$\,$\href{https://orcid.org/#1}{\textcolor{orcidlogocol}{\faOrcid}}}

\title{
\textbf{
Has Kronos devoured Planet Nine and its epigones?}
}

\author[]{
Lorenzo Iorio\,\orcid{0000-0003-4949-2694}
}

\affil[]{
\href{https://ror.org/01ehyh486}{Ministero dell' Istruzione e del Merito}
\\
Viale Unit\`{a} di Italia 68, I-70125, Bari, Italy \\ email: \href{mailto:lorenzo.iorio@libero.it}{\texttt{lorenzo.iorio@libero.it}}
}

\date{\today}

\providecommand{\keywords}[1]{keywords--- #1}

\begin{document}

\maketitle

\begin{center}
\begin{abstract}
\noindent
The Planet Nine hypothesis  encompasses a body of about 5-8 Earth's masses whose orbital plane would be inclined to the ecliptic by one or two tens of degrees and whose perihelion distance would be as large as about 240-385 astronomical units. Recently, a couple of his epigones have appeared: Planet X and Planet Y. The former is a sort of minor version of Planet Nine in that all its physical and orbital parameters would be smaller. Instead, the latter would have a mass ranging from that of Mercury to the Earth's one and semimajor axis within 100-200 astronomical units.
By using realistic upper bounds for the orbital precessions of Saturn, 
one can get insights on their position which, for Planet Nine, appears approximately confined around its aphelion. Planet Y can be just a Mercury-sized object at no less than about 125 astronomical units, while Planet X appears to be ruled out. 
Dedicated  data reductions by modeling such perturber(s) are required to check the present conclusions, to be intended as hints of what might be detectable should planetary ephemerides include them. A probe on the same route of Voyager 1 would be perturbed by Planet Nine by about 20-40 km after some decades.  
\end{abstract}
\end{center}

%
%
%
\section{Introduction}
In the last two and a half centuries or so, only two major planets have been discovered in our solar system compared to those known since ancient times: Uranus and Neptune. After the latter, none have been spotted in the last 180 years.
The intriguing possibility that another as yet undiscovered planet, generically referred to as Planet X, lurks in the distant outskirts of our solar system dates back even before the discovery of Neptune \citep{Grosser64}. Having resurfaced several times over the decades in astronomical research under various guises, its most recent version concerns mainly the impact of Planet X on the architecture of the Kuiper belt.

In this regards, the various observational clues about such a hypothesis  collected over the last twenty years have converged in 2016 when the hypothesis that a distant, still unseen major planetary body, dubbed provisionally Telisto\footnote{This name was proposed in \citet{2017Ap&SS.362...11I} from the ancient Greek word for \virg{farthest, most remote}  because of its supposed large heliocentric distance. \textcolor{black}{For a recent discussion on such a name, see \citet{2025Univ...11..405I}}.} or Planet Nine (P9),  may lurk in the remote peripheries of our solar system was put forth by \textcolor{black}{\citet{2014Natur.507..471T,2016AJ....152..221S} and, to a more compelling level,} by \citet{2016AJ....151...22B}. Such an object would explain the observed alignment of the lines of apsides and of the orbital angular momenta of just over ten distant Kuiper belt objects (KBOs) moving along eccentric orbits and not under the \textcolor{black}{strong} gravitational influence of Neptune \citep{2002Icar..157..269G,2003MNRAS.338..443E,2006Icar..184..589G}. On the other hand, no evidence for such an \textcolor{black}{alignment} was reported by \citet{2017AJ....153...33L,2017AJ....153...63S,2021PSJ.....2...59N}. Be that as it may, P9 would also be able to explain the existence of certain highly inclined trans-Neptunian objects (TNOs), whose orbital planes are nearly perpendicular to the ecliptic, and of some retrograde Centaurs exhibiting unusually high inclinations \citep{2016ApJ...833L...3B,2020CeMDA.132...44K}, and the tilt of the Sun's spin axis \citep{2017AJ....153...27G}. Recently, new lines of evidence supporting P9 were found in terms of the observed orbital properties of certain low-inclination asteroids orbiting inside Neptune's orbit \citep{2024ApJ...966L...8B}. For a recent review on P9, see \citep{2019PhR...805....1B}. The most recent version of this scenario envisages for such a hypothesized  heavy perturber a mass $m^{'}$ ranging from $4.9$ to $8.4$ Earth masses $m_\oplus$, a semimajor axis $a^{'}$ as large as 300 to 520 astronomical units (au), an inclination $I^{'}$ to the ecliptic of $11-21^\circ$, and a perihelion distance $q^{'}$ in the range $240-385$ au \citep{2021AJ....162..219B}, corresponding to  eccentricities $e^{'}\simeq 0.2-0.538$\textcolor{black}{, although most of them, in particular the \virg{best fit case}, have already been ruled out by \citet{2016A&A...587L...8F}}.  The best opportunity to discover P9 in the next few years appears to be the Vera C. Rubin Observatory just entered into service in Chile \citep{2024PhT....77k..22M}.

Recently, another distant planetary candidate, named once again in the long history of the proposed solar system's distant planets as Planet X (PX), was proposed \citep{2025ApJ...978..139S} to explain just the clustering of the longitudes of perihelia $\varpi$ of an extended sample of KBOs. It may be seen as a sort of \virg{reduced} or \virg{minor} version of P9 in the sense that its postulated physical and orbital parameters are systematically smaller than those of P9 itself, amounting to $m^{'}\simeq 3-5\,m_\oplus, a^{'}\simeq 260-320\,\mathrm{au}, I^{'}\simeq 2-12^\circ, q^{'}\simeq 150-270\,\mathrm{au}$. It is worth noting that P9 and this version of PX should be mutually exclusive since they would explain partially similar orbital clustering features in the Kuiper belt.

The latest addition to the hypothetical large family of still unseen solar system's planets postulated so far to explain some characteristics of the Kuiper belt's architecture seems to be Planet Y (PY) \citep{2025MNRAS.543L..27S}. With a mass $m^{'}$ between that of Mercury and the Earth, an inclination $I^{'}$ to the ecliptic of no less than $10^\circ$ and a semimajor axis $a^{'}$ between 100 and 200 au, it would explain the warping of the mean plane of the distant Kuiper belt.

In this paper, preliminary insights about the location of the aforementioned planetary candidates are tentatively obtained by means of the gravitational effects that they would unavoidably exert also on the known planets of the solar system perturbing their motion to a certain extent. In particular, the orbit of Saturn is adopted since it was recently constrained to a metre-level of accuracy \citep{2015IAUGA..2244873F,2020A&A...640A...7D} thanks to the long record of radio tracking data transmitted by the Cassini probe which explored the Kronian\footnote{From the Latin proper noun \textit{Cr\bo n\bo s, \li}, borrowed from the Greek proper noun $\qK\qr\aco\qn\qo\fs$, $\tn\qo\qu$, $\ho$ of one of the Titans born from Uranus and Gea, husband of Rhea and father, among others, of Zeus. He was identified with the Roman god Saturn. According to the myth, having been prophesied that one of his sons would supplant him and deprive him of power, he began to devour them one by one.} system for 13 years from 2004 to 2017 \citep{2019Sci...364.1046S}. As it will be shown later, they may not be considered as actual constraints, but just as hints of what might be detectable should such bodies were consistently searched for with planetary ephemerides purposely produced by including their gravitational tug as well. 
 
The paper is organized as follows. In Section\,\ref{Sec:2}, the analytical model of the planetary perturbations induced by a distant, pointlike object is presented and discussed. Section\,\ref{Sec:3} explains how to use the results obtained in Section\,\ref{Sec:2} in conjunction with the current upper bounds on the  orbital precessions of Saturn to explore the parameter space of a putative far planet. Such a task is implemented in Section\,\ref{Sec:4} for Planet Nine and its recent epigones by means of the orbit of Saturn assumed disturbed by them. The opportunities offered by a hypothetical improvement in our currently poor knowledge of Uranus's orbit are examined in Section\,\ref{Sec:5}. The possibility of using a dedicated deep-space probe is treated in  Section\,\ref{Sec:6} by using a Voyager 1-like path just for illustrative purposes. Section\,\ref{Sec:7} summarizes the findings and offers conclusions.
\section{The orbital perturbations induced by a distant, pointlike perturber}\label{Sec:2}
For the sake of clarity, the well known Keplerian orbital elements  will be employed in the ongoing analysis. They are as follows.
The semimajor axis $a$ and the eccentricity $0\leq e<1$ determine the size and the shape of the ellipse, respectively, in such a way that $e=0.0$ corresponds to a circular orbit. Often, the perihelion distance $q:=a\ton{1-e}$ is used instead of $e$.
The inclination $I$ is the tilt of the orbital plane to the reference plane $\qP$ adopted which, in the present case, is the mean ecliptic plane of the epoch J2000.0.
The longitude of the ascending node $\Omega$ is an angle in $\qP$ counted from the reference $x$ direction, which points toward the mean vernal equinox $\aries$ of the epoch J2000.0, to the intersection between $\qP$ and the orbital plane, known as line of nodes,  toward the ascending node $\ascnode$. The latter is the point where $\qP$ is crossed by the planet from below.
The argument of perihelion $\omega$ is an angle in the orbital plane counted from $\ascnode$ to the point of closest approach. 
The longitude of perihelion $\varpi$ is defined as $\varpi:=\Omega+\omega$; as such, it is a dogleg angle. \textcolor{black}{Finally, $\eta$ is the mean anomaly at epoch.}
Constant in a gravitationally bound two-body system, they instead change progressively over time if some other force acts on the otherwise unperturbed binary;
it is just the case of a distant perturbing body. If it is supposed to be far enough that its position does not appreciably vary over an orbital revolution of the disturbed planet, the long-term rates of change of the Keplerian orbital elements of the latter can be analytically calculated with the standard perturbative techniques to the Newtonian quadrupolar order of the additional potential due to $m^{'}$ \citep{1991AJ....101.2274H}. 
%
%
%
They turn out to be
\textcolor{black}{
\begin{align}
\ang{\dert a t} \lb{dadt} & = 0, \acap
\ang{\dert e t} \lb{dedt} & = \rp{15 e\upmu^{'}\sqrt{1 - e^2}}{4{r^{'}}^3 n_\mathrm{K}}\qua{-2 \mathtt{Rl}\,\mathtt{Rm} \cos 2\omega + \ton{\mathtt{Rl}^2 - \mathtt{Rm}^2} \sin 2\omega}, \acap
\ang{\dert I t} \lb{dIdt} & = \rp{3\upmu^{'}\mathtt{Rh}}{4{r^{'}}^3 n_\mathrm{K}\sqrt{1 - e^2}}\qua{\ton{2 + 3 e^2} \mathtt{Rl} + 5 e^2 \ton{\mathtt{Rl}\cos 2\omega + \mathtt{Rm}\sin 2\omega}}, \acap
\ang{\dert{\mathit{\Omega}} t} \lb{dOdt} & = \rp{3\upmu^{'}\csc I \mathtt{Rh}}{4{r^{'}}^3 n_\mathrm{K}\sqrt{1 - e^2}}\grf{\ton{2 + 3 e^2} \mathtt{Rm} +  5 e^2\ton{-\mathtt{Rm}\cos 2\omega + \mathtt{Rl}\sin 2\omega}}, \acap
\ang{\dert \omega t} \nonumber & = -\rp{3\upmu^{'}}{4{r^{'}}^3 n_\mathrm{K}\sqrt{1 - e^2}}\grf{\ton{-1 + e^2} \qua{-2 + 3\ton{\mathtt{Rl}^2 + \mathtt{Rm}^2}} + \ton{2 + 3 e^2} \mathtt{Rh}\,\mathtt{Rm} \cot I  \right.\acap
\lb{dodt}&\left. -5\qua{-\ton{-1 + e^2}\ton{\mathtt{Rl}^2 - \mathtt{Rm}^2} + e^2 \mathtt{Rh}\,\mathtt{Rm} \cot I} \cos 2\omega + 5 \mathtt{Rl} \qua{2 \mathtt{Rm} \ton{-1 + e^2} + e^2 \mathtt{Rh} \cot I} \sin 2\omega}, \acap
\ang{\dert \eta t} \nonumber & = -\rp{\upmu^{'}}{4{r^{'}}^3 n_\mathrm{K}}\grf{\ton{7 + 3 e^2} \qua{-2 + 3\ton{\mathtt{Rl}^2 + \mathtt{Rm}^2}} + 15 \ton{1 + e^2}\ton{\mathtt{Rl}^2 - \mathtt{Rm}^2}\cos 2\omega + \right.\acap
\lb{detadt} & \left. + 30 \ton{1 + e^2} \mathtt{Rl}\,\mathtt{Rm}\sin 2\omega}.
\end{align}
From \rfrs{dOdt}{dodt}, the equation for the precession of the longitude of the perihelion
\eqi
\dert\varpi t = \dert{\mathit{\Omega}}t + \dert\omega t\lb{dvarpidt}
\eqf
is straightforwardly obtained.
}
%
%
%
%
%
%
%
%
%

\textcolor{black}{In \rfrs{dadt}{detadt}, the Keplerian mean motion of the perturbed planet is $\nk:=\sqrt{\upmu_\odot/a^3}$,
where $\upmu_\odot:=GM_\odot$ is the Sun's standard gravitational parameter given by the product of the Newtonian constant of gravitation $G$ by its mass $M_\odot$. 
In \rfrs{dadt}{detadt}, $\upmu^{'}:=Gm^{'}$ is the standard gravitational parameter of the distant perturber, and $r^{'}$  is its heliocentric distance given by
\eqi
r^{'} = \rp{a^{'}\ton{1 - {e^{'}}^2}}{1 + e^{'}\cos f^{'}},
\eqf
where $e^{'}$ and $f^{'}$ are its eccentricity and its true anomaly, respectively.
Moreover, in \rfrs{dadt}{detadt}, $\mathtt{Rl},\mathtt{Rm},\mathtt{Rh}$ are defined as
\begin{align}
\mathtt{Rl} \lb{Rl}& :={\bds{\hat{r}}}^{'}\bds\cdot\bds{\hat{l}}, \acap
\mathtt{Rm} \lb{Rm}& :={\bds{\hat{r}}}^{'}\bds\cdot\bds{\hat{m}}, \acap
\mathtt{Rh} \lb{Rh}& :={\bds{\hat{r}}}^{'}\bds\cdot\bds{\hat{h}},
\end{align}
where
\begin{align}
\bds{\hat{l}} \lb{elle} & = \grf{\cos\Omega, \sin\Omega, 0},\acap
\bds{\hat{m}} \lb{emme} & = \grf{-\cos I\sin\Omega, \cos I\cos\Omega, \sin I},\acap
\bds{\hat{h}} \lb{acca} & = \grf{\sin I\sin\Omega, -\sin I\cos\Omega, \cos I}.
\end{align}
While the unit vector $\bds{\hat{l}}$ is directed along the line of nodes toward the ascending node $\ascnode$, $\bds{\hat{h}}$ is aligned with the orbital angular momentum. Finally, $\bds{\hat{m}}$, lying in the orbital plane, is defined such that the relation $\bds{\hat{l}}\bds\times \bds{\hat{m}}= \bds{\hat{h}}$ holds.
The unit vector ${\bds{\hat{r}}}^{'}$ of the position vector ${\bds r}^{'}$ of the distant perturber entering \rfrs{Rl}{Rh} can be expressed as
\eqi
{\bds{\hat{r}}}^{'} = {\bds{\hat{l}}}^{'}\cos u^{'} + {\bds{\hat{m}}}^{'}\sin u^{'},\lb{rY}
\eqf
where the unit vectors
\begin{align}
{\bds{\hat{l}}}^{'} \lb{elleY} & = \grf{\cos\Omega^{'}, \sin\Omega^{'}, 0},\acap
{\bds{\hat{m}}}^{'} \lb{emmeY} & = \grf{-\cos I^{'}\sin\Omega^{'}, \cos I^{'}\cos\Omega^{'}, \sin I^{'}}
\end{align}
span its orbital plane, and $u^{'}:=\omega^{'} + f^{'}$ is its argument of latitude.
Over one orbital period $P_\mathrm{K}=2\pi/\nk$ of the perturbed planet, \rfrs{rY}{emmeY}, which implicitly depend on time $t$ essentially through $f^{'}$, can be considered constant.}
\section{Using the orbital precessions of a disturbed planet to constrain the position of a distant perturber}\label{Sec:3}
Let $\upkappa$ be any of the Keplerian orbital elements of some known major planet of the solar system. Any anomalous rate of change of that orbital element, averaged over one orbital revolution of the planet under consideration, with respect to standard dynamics modelled in the softwares used to produce planetary ephemerides, is  dubbed as $\Delta\dot\upkappa$. Furthermore, let $\upsigma_{\dot\upkappa}$ be an observationally inferred measure of the uncertainty in the rate of variation of $\upkappa$ determined over a certain time span. Since, to date, no secular variations $\Delta\dot\upkappa$ have been determined that are different from zero at a statistically significant level, it is possible to write 
\eqi
\left|\Delta\dot\upkappa\right|\leq\sigma_{\dot\upkappa}\lb{condiz}
\eqf
for all of them.

At present, the most recent assessment of the uncertainties $\upsigma_{\dot\upkappa}$ in the long term rates of change of  $a,e,I,\Omega,\varpi$ for all the known planets of the solar system and also for Pluto dates back to 2019 \citep{2019AJ....157..220I}. They were calculated from the published formal, statistical errors in the non-singular orbital elements of the solar system's planets determined by \citet{2018AstL...44..554P} with the EPM2017 ephemerides, belonging to the series of Ephemeris\footnote{To date, the most renown planetary ephemerides available other than those of the EPM series are the planetary and lunar ephemerides of the series Development Ephemerides (DE/LE) produced by the Jet Propulsion Laboratory (JPL) of the National Aeronautics and Space Administration (NASA), and the planetary ephemerides of the series Intégrateur Numérique Planétaire de l'Observatoire de Paris (INPOP) produced by the past Institut de mécanique céleste et de calcul des éphémérides (IMCCE).} of Planets and the Moon (EPM) developed at the Institute of Applied Astronomy (IAA) of the Russian Academy of Sciences (RAS), without modeling any hypothetical distant planet. 

In principle, by straightforwardly identifying $\Delta\dot\upkappa$ with the analytically calculated long-term rates of change of the Keplerian orbital elements of a planet due to the action of a putative distant perturber, considered as functions of $f^\mathrm{'},\Omega^\mathrm{'},\omega^\mathrm{'}$ for any given values of $a^\mathrm{'},e^\mathrm{'},I^\mathrm{'}$, it is possible to infer preliminary insights on that part of the parameter space pertaining its position in the space. Since uncertainties for all the orbital elements are available \citep{2019AJ....157..220I}, it is possible to obtain tighter results by imposing that all the rates of change of, say, Saturn simultaneously fulfil the condition of \rfr{condiz}. The choice of the ringed gaseous giant is motivated by the fact that if on the one hand Uranus, Neptune and Pluto, being much further from the Sun, are, in principle, more sensitive to hypothetical detached perturber, on the other hand their orbital precessions are known with a much greater uncertainty than Saturn's, as shown in Section \ref{Sec:5}. 
The latter ones are displayed in Table \ref{Tab:1}.
\begin{table}[ht]
\centering
\caption{Formal uncertainties, in milliarcseconds per century (mas cty$^{-1}$), of the long-term rates of change of the eccentricity $e$, the inclination $I$, the longitude of the ascending node $\Omega$ and the longitude of perihelion $\varpi$ of Saturn according to Table 1 of \citet{2019AJ....157..220I}, based on Table 3 of \citet{2018AstL...44..554P}.
}\lb{Tab:1}
\vspace{0.3cm}
\begin{tabular}{l l l l}
\toprule
$\upsigma_{\dot e}$ & $\upsigma_{\dot I}$ & $\upsigma_{\dot\Omega}$ & $\upsigma_{\dot\varpi}$ \\
\midrule
$0.0023$ & $0.063$ & $1.806$ & $0.067$  \\
\bottomrule
\end{tabular}
\end{table}
They could be deemed not so far from representing rather realistic accuracies for Saturn. This can be understood as follows. First, 
the uncertainty in the perihelion can be translated into a linear accuracy of the order of\footnote{Here, the impact of the error in the eccentricity is neglected since it is about 30 times smaller than that on the perihelion.} $\upsigma_r\simeq ae\upsigma_{\Delta\varpi}$. The value of $\upsigma_{\dot\varpi}$ in the last column of Table \ref{Tab:1} yields $\upsigma_r\simeq 3.4\,\mathrm{m}$ over 13 years, which agrees with the published Cassini-based orbital accuracy \citep{2020A&A...640A...7D,2021AJ....161..105P}; in particular, Figure 11 of \citet{2021AJ....161..105P} shows that the root-mean-square (rms) residuals of the Cassini range data against the ephemerides DE440 is about 3 m over the same time span. Another hint for the linear accuracy in the normal direction to the Kronian orbit may be evaluated with the aid of $\upsigma_{\dot I},\upsigma_{\dot\Omega}$ in Table \ref{Tab:1} to be of the order of 180 m over the same time span. Furthermore,  for Saturn, $0.067$ mas cty$^{-1}$ corresponds to an uncertainty of the order of $5\times 10^{-3}$ in the parameterized post-Newtonian (PPN) parameters $\beta$ and $\gamma$ in terms of which the general relativistic precession is expressed as
\eqi
\dot\varpi_\mathrm{GR} =\ton{\rp{2+2\gamma-\beta}{3}}\rp{3\nk\upmu}{c^2 a\ton{1-e^2}},
\eqf
where $c$ is the speed of light in vacuum, in agreement with the discrepancies between different ephemerides for the Kronian heliocentric distance \citep{2021NSTIM.110.....F}.
Anyway, \citet{2018AstL...44..554P} warned that the actual uncertainties may be up to one
order of magnitude larger. Caution is in order also because a more accurate procedure would imply the use of dedicated ephemerides purposely built by explicitly modeling the putative distant perturber, as done in \citet{2016A&A...587L...8F,2016AJ....152...80H,2016AJ....152...94H,2020A&A...640A...6F}; see also \citep{2023PSJ.....4...66G}. Otherwise, its signal, if real, may, in principle, be partially or totally removed, being absorbed in values of the solved-for parameters which are estimated in the data reduction procedure, thus resulting in too optimistically tight constraints \citep{Fienga24}. On the other hand, even if P9 were explicitly modeled and a specific data reduction performed, it could still happen that some other dynamical effect, whether standard or not, has been neglected or is poorly modeled. In this case, the parameters estimated in these specific ephemerides could still be biased by these other unmodeled or poorly modeled forces. Indeed, it must be kept in mind that a choice of parameters to estimate is always made. It should also be said that the approach followed here is widely adopted by most researchers worldwide to put constraints, e.g., on many non-standard dynamical effects such as those predicted by a host of modified gravity models. Indeed, considering that the production of ephemerides is an extremely laborious and time-consuming task that can only be completed by highly specialized scientists
who are not always, if ever, interested in hypothetical exotic effects of various kinds, it would be unthinkable to believe that for each of these features one could reprocess decades of observations each time, purposely modeling and estimating the dynamical effect of interest. Be that as it may, in a previous analysis \citep{2017Ap&SS.362...11I} the present method did not provide constraints in contrast to those inferred by reprocessing the data record by including P9 in the dynamical force models \citep{2016A&A...587L...8F}.
%
\section{Tentative constraints on Planet Y, Planet X and Planet Nine}\label{Sec:4}
Let one simultaneously impose that the absolute values of the rates of changes of all the Kronian orbital elements induced by a distant perturber are smaller than the formal uncertainties listed in Table \ref{Tab:1}. 
In the case of PY, it turns out that no allowed regions exist at all in the $\{f_\mathrm{Y},\Omega_\mathrm{Y},\omega_\mathrm{Y}\}$ parameter space for either an Earth-mass body or a Mercury-sized one with the orbital characteristics envisaged for it by the PY hypothesis.

\begin{figure}[ht!]
\centering
\begin{tabular}{cc}
\includegraphics[width = 6.5 cm]{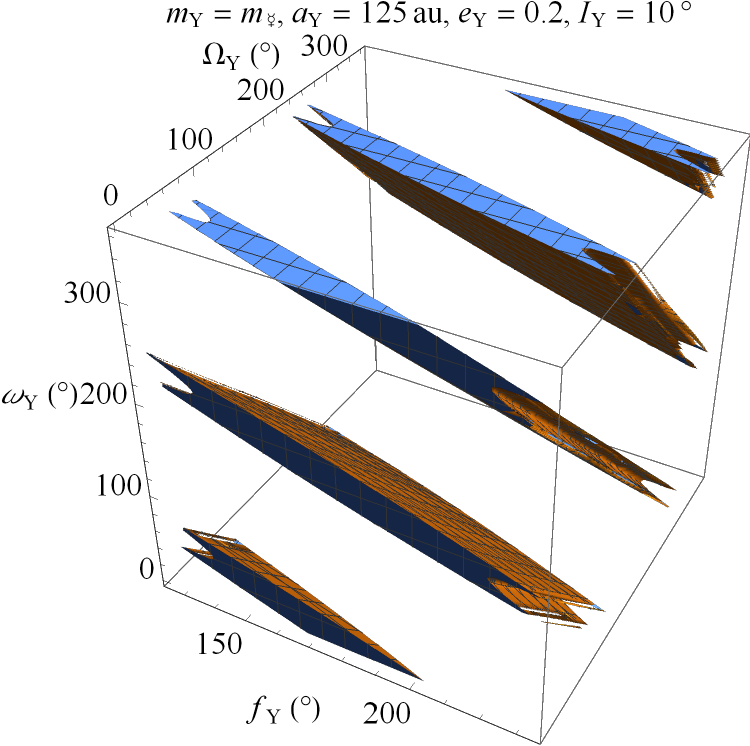} & \includegraphics[width = 6.5 cm]{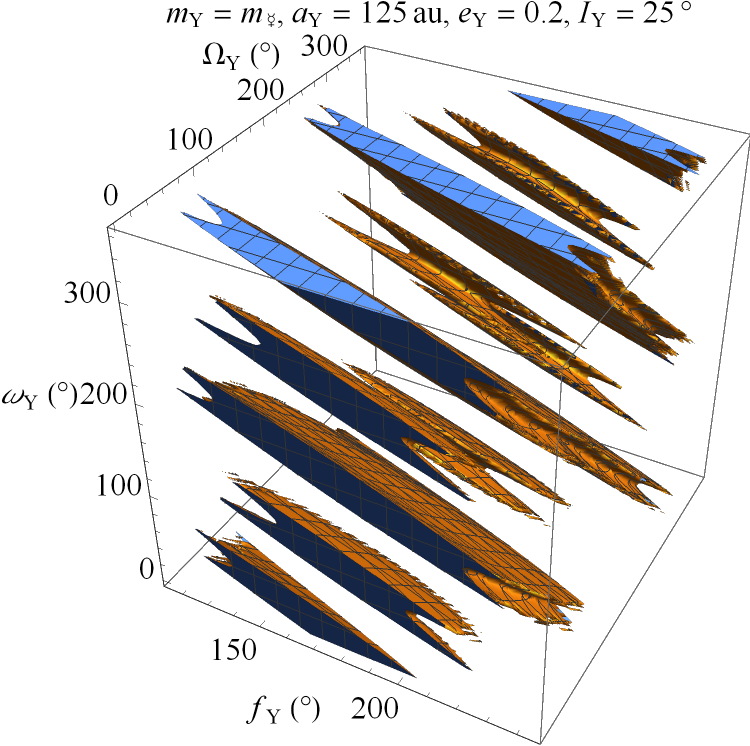}\\
\includegraphics[width = 6.5 cm]{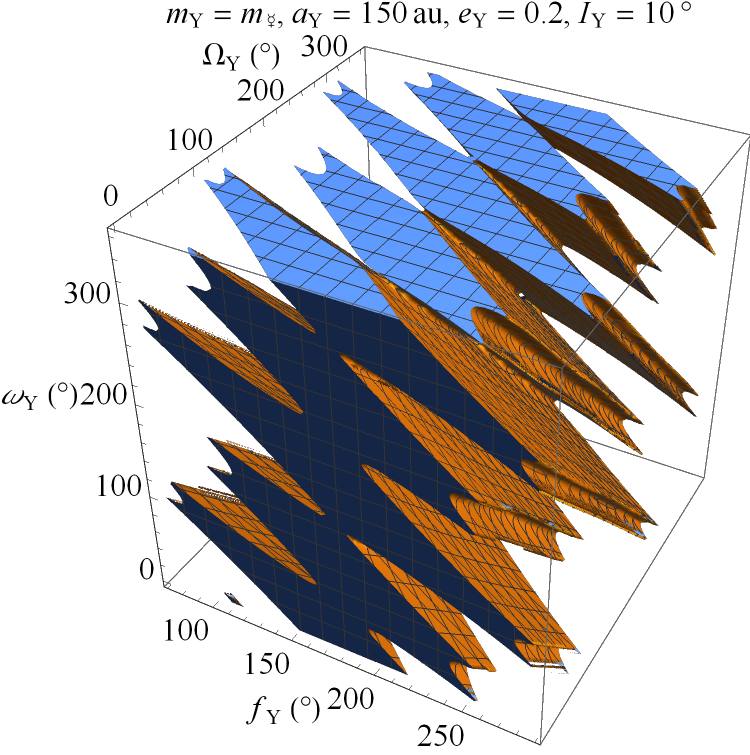} & \includegraphics[width = 6.5 cm]{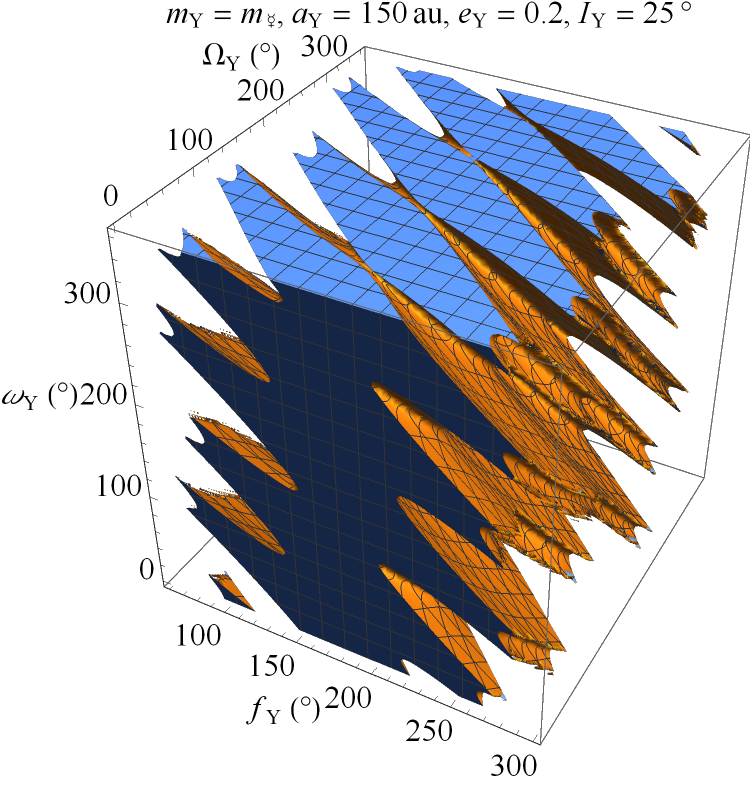}\\
\includegraphics[width = 6.5 cm]{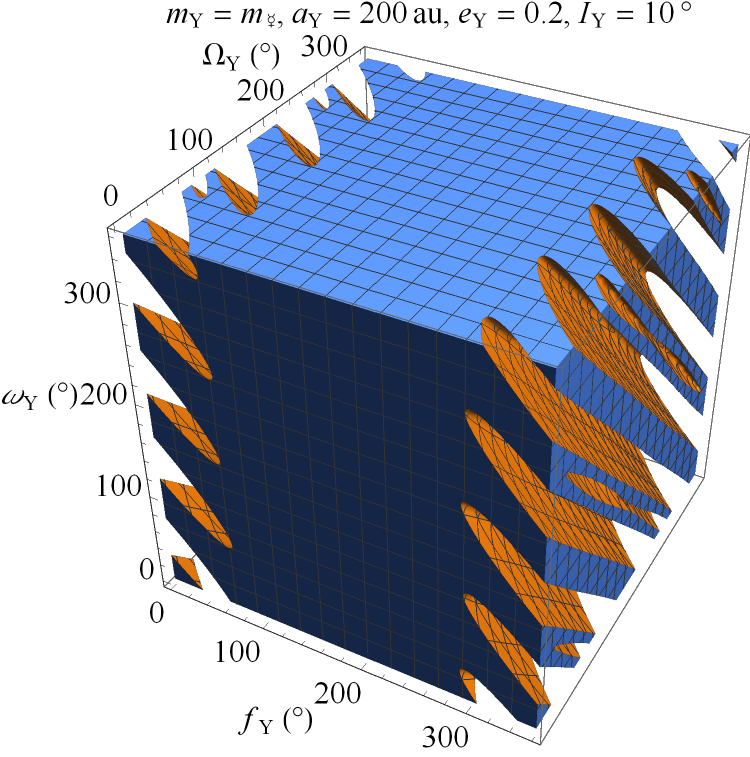} & \includegraphics[width = 6.5 cm]{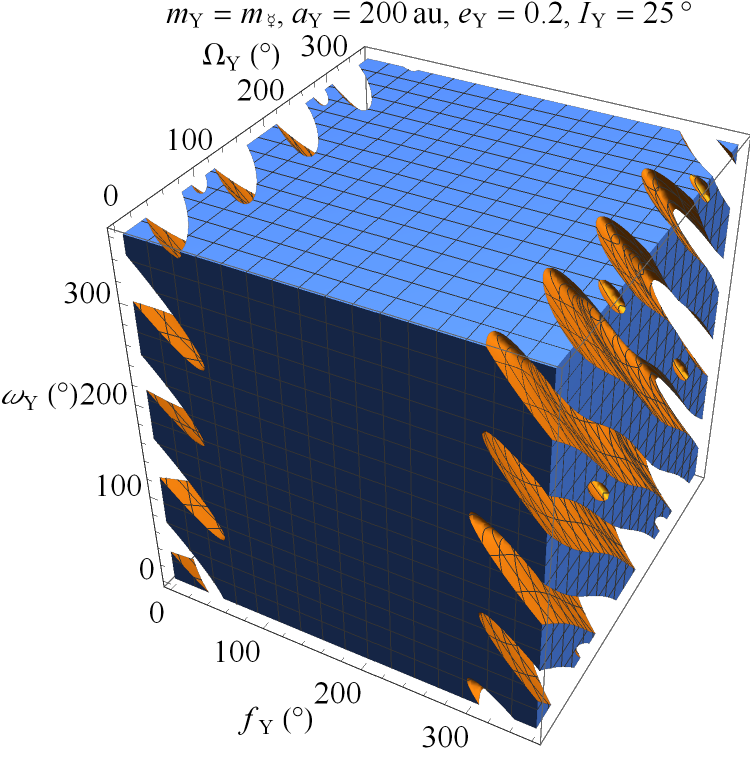}\\
\end{tabular}
\caption{Allowed regions in the $\grf{f_\mathrm{Y},\Omega_\mathrm{Y},\omega_\mathrm{Y}}$ parameter space for an elliptical orbit of a Mercury-mass Planet Y characterized by $e_\mathrm{Y}=0.2$ and different values of the semimajor axis $a_\mathrm{Y}$ (125 au, 150 au and 200 au) and the inclination $I_\mathrm{Y}$ (10$^\circ$ and 25$^\circ$) to the ecliptic. They were  inferred by imposing  that the theoretical rates of change of $e,I,\Omega,\varpi$ of Saturn due to Planet Y, calculated with \rfrs{dadt}{dvarpidt}, simultaneously fulfil the condition of \rfr{condiz}
by using the uncertainties in Table \ref{Tab:1} rescaled by a factor of ten.}\label{Fig:1}
\end{figure}

If the uncertainties $\upsigma_{\dot\upkappa}$ in Table \ref{Tab:1} are conservatively rescaled by a factor of ten, only the scenario with $m_\mathrm{Y}=m_{\mercury},\,a_\mathrm{Y}\gtrsim 125\,\mathrm{au}$ is not discarded. It looks less constrained in the $\{f_\mathrm{Y},\Omega_\mathrm{Y},\omega_\mathrm{Y}\}$ parameter space the further away the planet is, as shown by Figure \ref{Fig:1}. In producing them, a moderate value of the eccentricity ($e_\mathrm{Y}=0.2$) is realistically assumed. Indeed, should it be extremely eccentric, it would produce an apsidal clustering of TNOs distinct from that by PX/P9, which, however, is not observed. On the other hand, it is unlikely that PY may have a completely circular orbit because of its assumed process of formation\footnote{It would have been scattered from the region where the known planets exist by other planetary embryos in the remote past, which should have left PY with some residual eccentricity.} \citep{2025MNRAS.543L..27S}.
%
%
%
%
%
%
%
%
%

The PX  scenario is completely ruled out by the combined orbital precessions of Saturn, even by rescaling their formal uncertainties listed in Table \ref{Tab:1} by a factor of 10. 
%
%
%
%

\begin{figure}[ht!]
\centering
\begin{tabular}{cc}
\includegraphics[width = 6.5 cm]{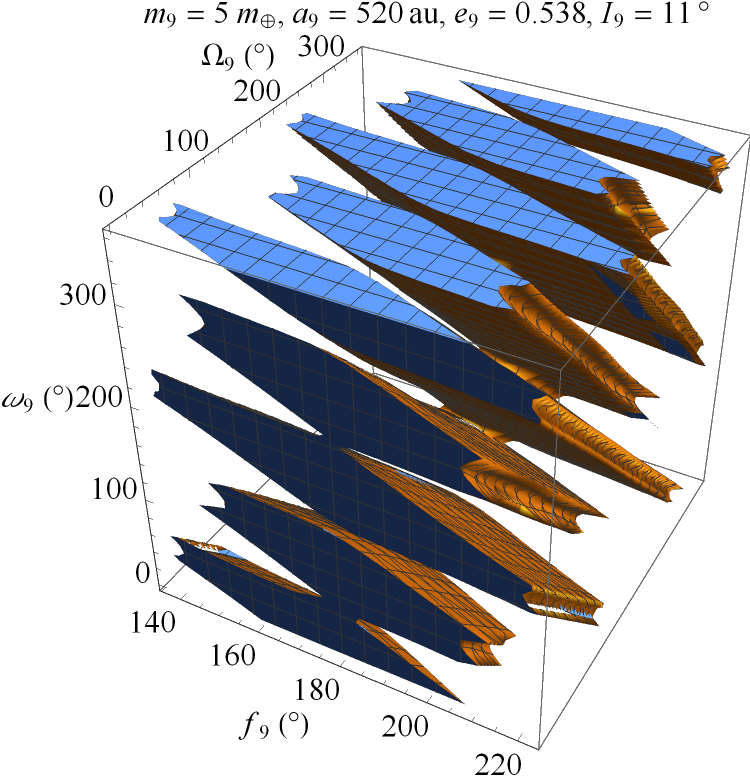} & \includegraphics[width = 6.5 cm]{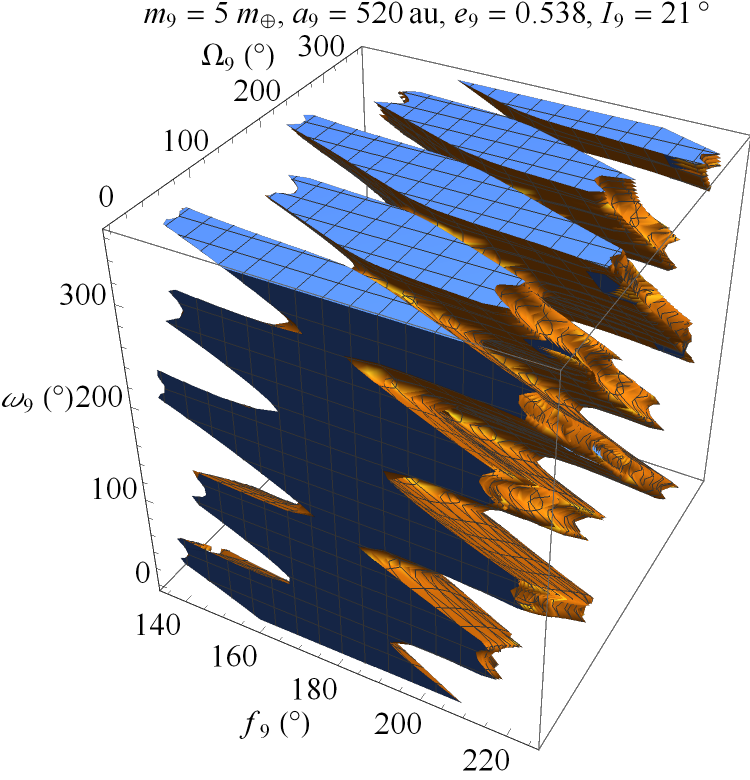}\\
\includegraphics[width = 6.5 cm]{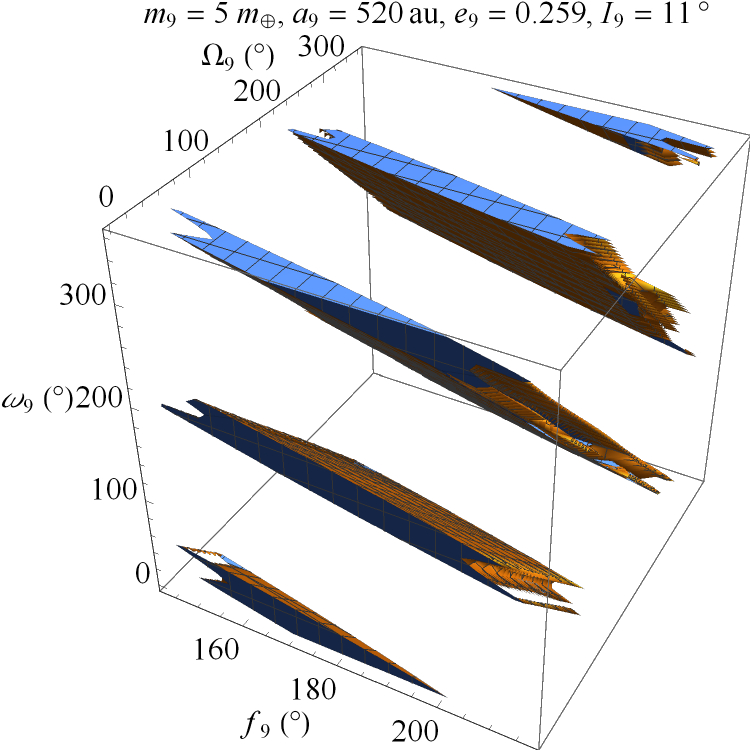} & \includegraphics[width = 6.5 cm]{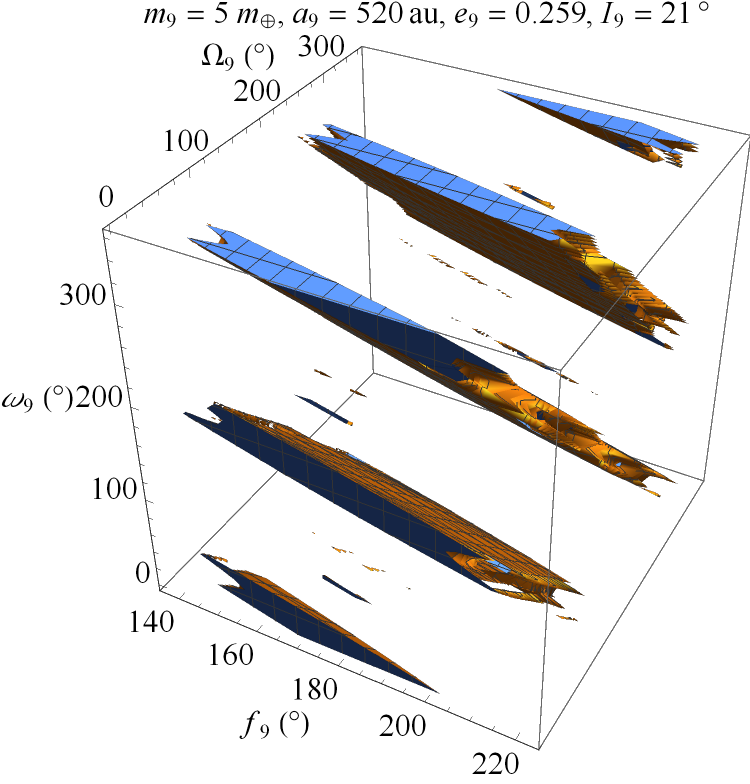}\\
\end{tabular}
\caption{Allowed regions in the $\grf{f_9,\Omega_9,\omega_9}$ parameter space for an elliptical orbit of a $5\,m_\oplus$ Planet Nine characterized by $a_9=520\,\mathrm{au}$ and different values of the eccentricity $e_9$ (0.259 and 0.538) and the inclination $I_9$ (11$^\circ$ and 21$^\circ$) to the ecliptic. They were  inferred by imposing  that the theoretical rates of change of $e,I,\Omega,\varpi$ of Saturn due to Planet Nine, calculated with \rfrs{dadt}{dvarpidt}, simultaneously fulfil the condition of \rfr{condiz} by using the uncertainties in Table \ref{Tab:1} rescaled by a factor of ten.}\label{Fig:2}
\end{figure}

\begin{figure}[ht!]
\centering
\begin{tabular}{cc}
\includegraphics[width = 6.5 cm]{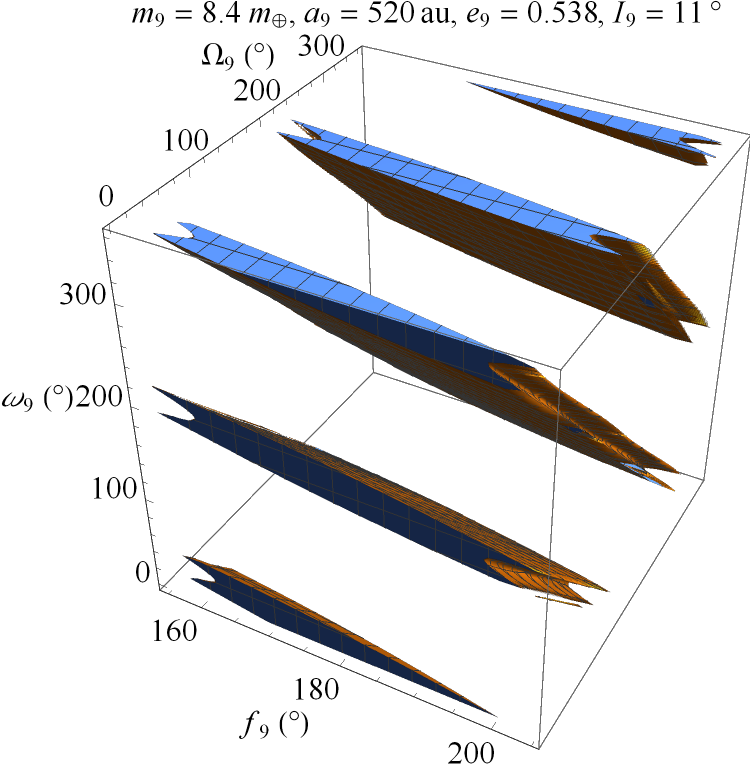} & \includegraphics[width = 6.5 cm]{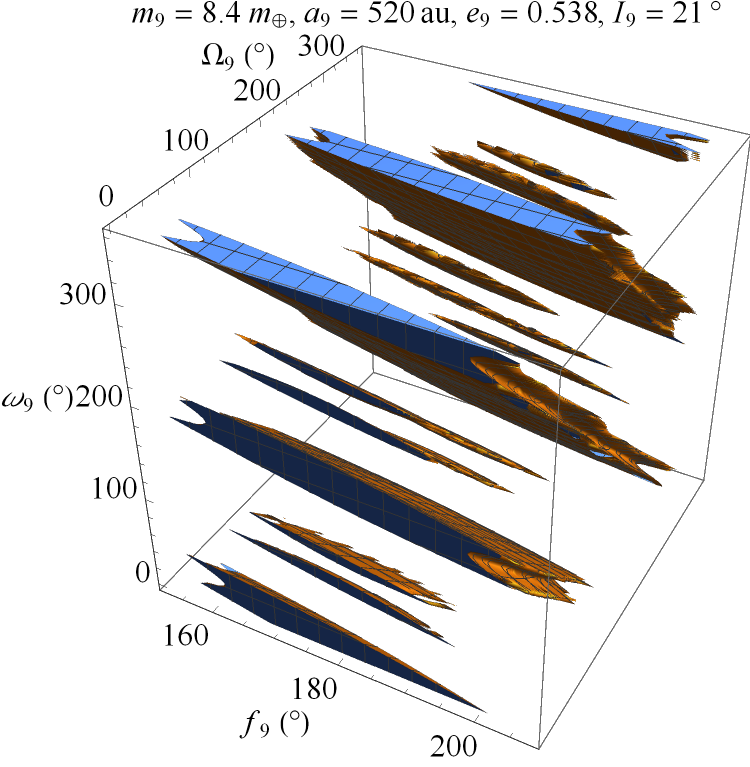}\\
\end{tabular}
\caption{Allowed regions in the $\grf{f_9,\Omega_9,\omega_9}$ parameter space for an elliptical orbit of a $8.4\,m_\oplus$ Planet Nine characterized by $a_9=520\,\mathrm{au}, e_9=0.538$ and different values of the inclination $I_9$ (11$^\circ$ and 21$^\circ$) to the ecliptic. They were  inferred by imposing  that the theoretical rates of change of $e,I,\Omega,\varpi$ of Saturn due to Planet Nine, calculated with \rfrs{dadt}{dvarpidt}, simultaneously fulfil the condition of \rfr{condiz} by using the uncertainties in Table \ref{Tab:1} rescaled by a factor of ten.}\label{Fig:3}
\end{figure}

\begin{figure}[ht!]
\centering
\begin{tabular}{cc}
\includegraphics[width = 5.5 cm]{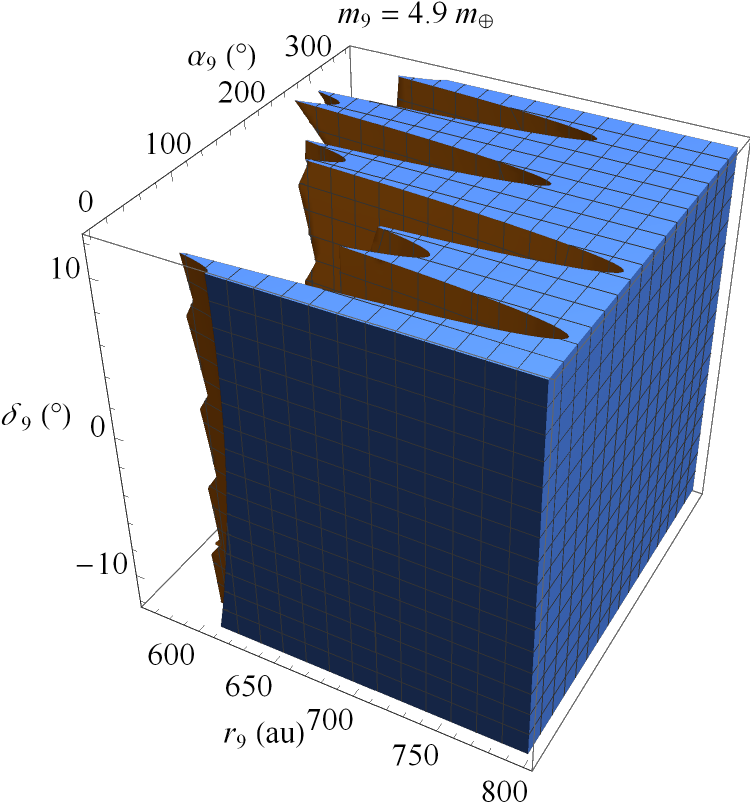} & \includegraphics[width = 5.5 cm]{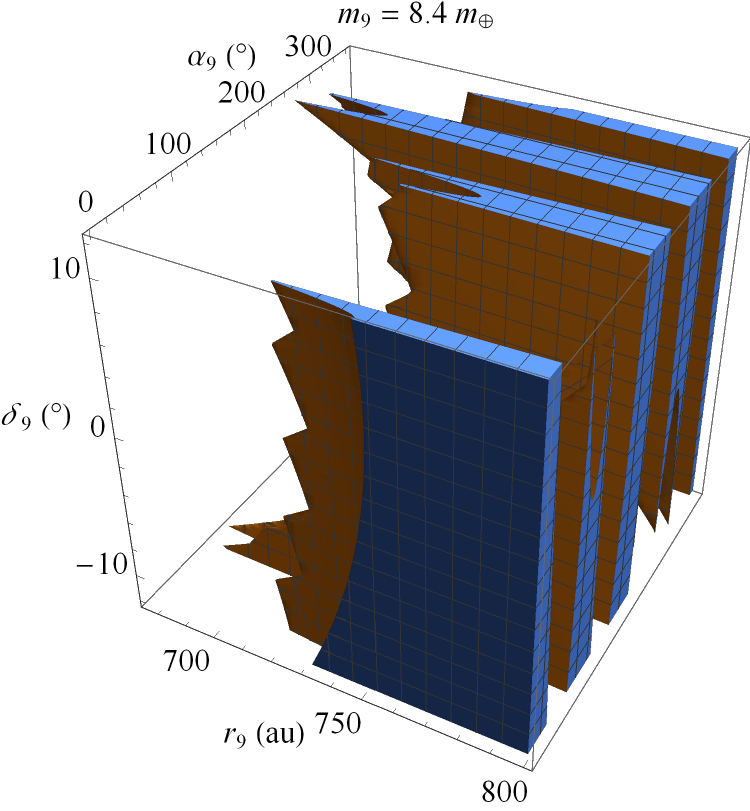}\\
\includegraphics[width = 5.5 cm]{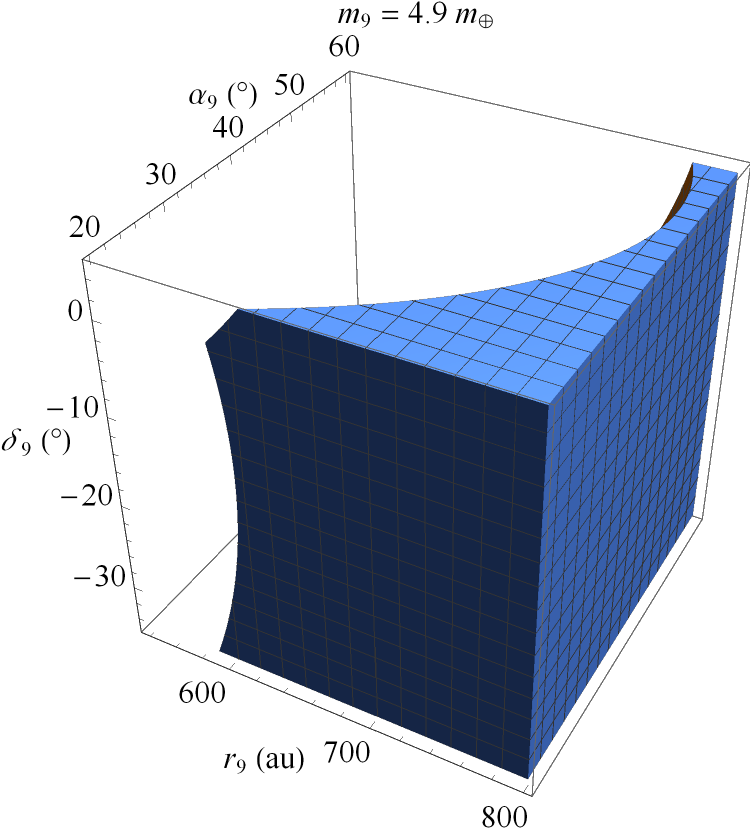} & \includegraphics[width = 5.5 cm]{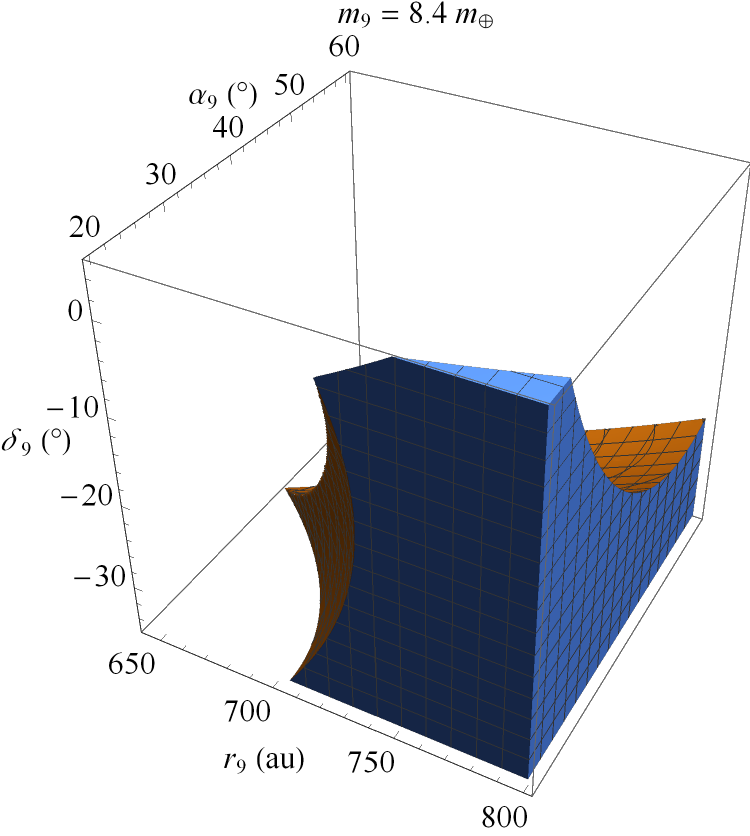}\\
\includegraphics[width = 5.5 cm]{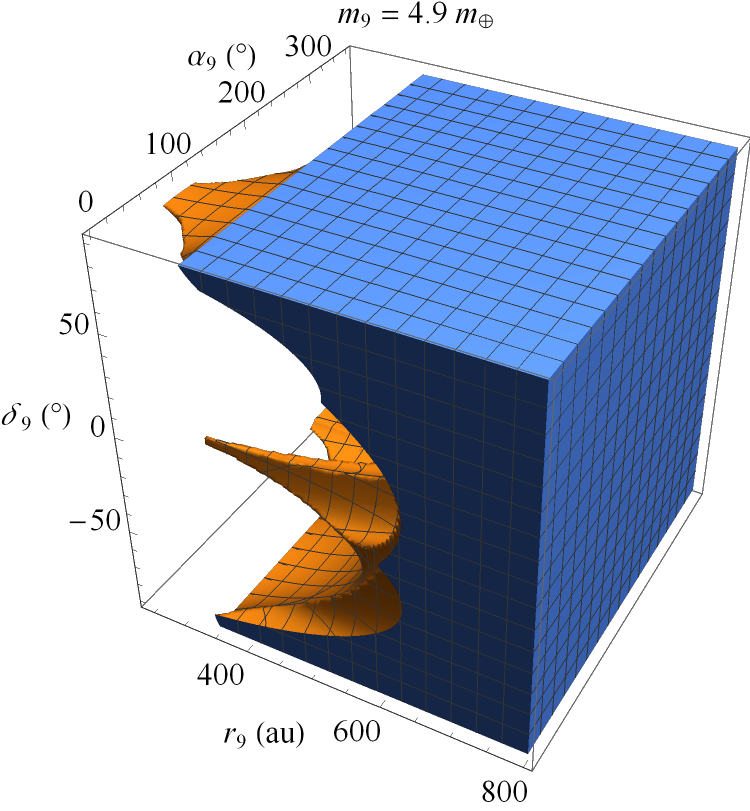} & \includegraphics[width = 5.5 cm]{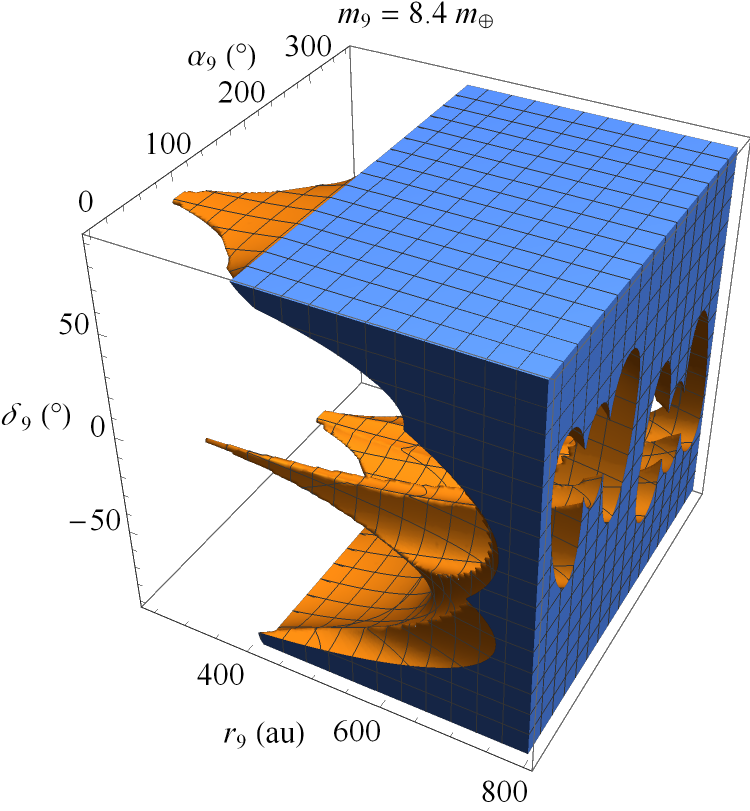}\\
\end{tabular}
\caption{Allowed regions in the $\grf{r_9,\alpha_9,\delta_9}$ parameter space for an elliptical orbit of Planet Nine characterized by different values of its mass $m_9$ ($4.9\,m_\oplus$ and $8.4\,m_\oplus$). They were  inferred by imposing  that the theoretical Kronian rates of change of $e,I,\Omega,\varpi$, referred to the International Celestial Reference Frame (ICRF), due to Planet Nine, calculated with \rfrs{dadt}{dvarpidt}, simultaneously fulfil the condition of \rfr{condiz} by using the uncertainties in Table \ref{Tab:1} rescaled by a factor of ten. The ranges of variation for the heliocentric distance $r_9$ and the astrometric angles $\alpha_9,\delta_9$ of P9 are retrieved from \citet{2021AJ....162..219B} (upper row) and \citet{2016AJ....152...94H} (middle row), while the lower row shows the allowed regions without assuming any a priori bounds for $\alpha_9,\delta_9$.}\label{Fig:4}
\end{figure}

As far as P9 is concerned, there is room for its existence mainly around its aphelion, as shown by Figure \ref{Fig:2} ($m_9=5\,m_\oplus$) and Figure \ref{Fig:3} ($m_9=8.4\,m_\oplus$) obtained with the rescaled values of Table \ref{Tab:1}.
In particular, for a 5 Earth mass body, allowed regions exist for $a_9=520\,\mathrm{au}$. While $\Omega_9$ and $\omega_9$ can take almost any value from $0^\circ$ to 360$^\circ$ for $e_9=0.538$, they appear more constrained within the full angular range if $e_9=0.259$, being confined in disjointed \virg{islands} in the $\grf{f_9,\Omega_9,\omega_9}$ parameter space; in both cases, $f_9$ stays approximately  within the range $140^\circ-220^\circ$. The heaviest version of P9 ($m_9=8.4\,m_\oplus$) appears limited to $a_9=520\,\mathrm{au},\,e_9=0.538$. In this case, only well detached allowed regions in the parameter space exist, corresponding to few values of $\Omega_9$ and $\omega_9$ in the full angular range, and the true anomaly $f_9$ is tighter constrained within about $160^\circ-200^\circ$. 
In Figure \ref{Fig:4}, the allowed regions in the parameter space $\grf{r_9,\alpha_9,\delta_9}$ are depicted, where $r_9$ is the heliocentric distance of P9, and $\alpha_9,\delta_9$ are its right ascension (RA) and declination (decl.), respectively. According to \citet{2021AJ....162..219B}, while the RA is unconstrained, the decl. would range from $-12^\circ$ to $12^\circ$. Instead, \citet{2016AJ....152...94H} found $20^\circ\leq \alpha_9 \leq 60^\circ,-35^\circ\leq\delta_9\leq +5^\circ$. As far as the maximum heliocentric distance envisaged by \citet{2021AJ....162..219B} is concerned, $a_9=520\,\mathrm{au}$ and $q_9=240\,\mathrm{au}$ yield an aphelion distance as large as $Q_9=800\,\mathrm{au}$.
It can be noted that, within the expected interval for its declination by \citet{2021AJ....162..219B}, the lighter version of P9 can exist at not less than about 560 au for just a narrow range of values of its RA, and at not less than about 600 au for most of the values of its RA. Instead, a $8.4\,m_\oplus$ body has its RA constrained to a few disjoint allowed intervals, while its heliocentric distance cannot be smaller than about 670 au. If the ranges for $\alpha_9,\delta_9$ by \citet{2016AJ....152...94H} are taken into account,  the allowed regions with $\alpha_9$ close to $20^\circ$ are generally more extended than those characterized by larger values of the RA: they are bounded from below at 600 au ($m_9 = 4.9\,m_\oplus$) and 700 au ($m_9 = 8.4\,m_\oplus$). Anyway, they are overall rather small. Figure \ref{Fig:4} depicts also the allowed regions by removing the previous limitations on $\alpha_9,\delta_9$. In general, a body at highest \textcolor{black}{absolute} value of declination may exist at shorter heliocentric distances, even smaller than 400 au.
All such findings are, essentially, not in contrast with those by \citet{2020A&A...640A...6F} inferred with a version of the INPOP19a ephemerides purposely modified to include the attraction of P9 as well, as in the case of \citet{2017Ap&SS.362...11I}. Indeed, \citet{2020A&A...640A...6F} concluded that a body with 5 Earth masses cannot be closer than 500 au. Moreover, they were not able to indicate a precise position for P9, providing allowed zones for P9 where its existence is compatible with the accuracy of the INPOP planetary ephemerides. Also \citet{2016AJ....152...94H} were not able to spot a unique position of P9 in space, but only a set of allowed regions in the $\grf{\alpha_9,\delta_9}$ parameter space.
%
\section{What if Uranus's orbit was known as accurately as Saturn's?}\label{Sec:5}
In principle, the orbit of Uranus, whose orbital period is 84 yr long since it is 20 au away from the Sun, would be more affected than Saturn's by a distant planetary body like those considered so far. Unfortunately, at present, we can only rely on astrometric optical observations of the bluish planet, apart from very few radiometric data collected by the Voyager 2 probe during its flyby in 1986 \citep{1986Sci...233...39S,1987JGR....9214873S}. Thus, its orbital rates of change are currently constrained at a very poor level, as shown in Table \ref{Tab:2}.
\begin{table}[ht]
\centering
\caption{Formal uncertainties, in milliarcseconds per century (mas cty$^{-1}$), of the long-term rates of change of the eccentricity $e$, the inclination $I$, the longitude of the ascending node $\Omega$ and the longitude of perihelion $\varpi$ of Uranus according to Table 1 of \citet{2019AJ....157..220I}, based on Table 3 of \citet{2018AstL...44..554P}.
}\lb{Tab:2}
\vspace{0.3cm}
\begin{tabular}{l l l l}
\toprule
$\upsigma_{\dot e}$ & $\upsigma_{\dot I}$ & $\upsigma_{\dot\Omega}$ & $\upsigma_{\dot\varpi}$ \\
\midrule
$2.732$ & $3.827$ & $269.177$ & $47.998$  \\
\bottomrule
\end{tabular}
\end{table}
Such large formal errors, which are likely to be rescaled by at least a factor of ten, make it pointless to repeat the analysis of Section \ref{Sec:4} because it turns out that the entire parameter space would be allowed for the previously considered planetary candidates. Suffice it to say that $\upsigma_{\dot\varpi}$ in Table \ref{Tab:2} corresponds to a formal linear uncertainty as large as $\upsigma_r\simeq a e\upsigma_{\dot\varpi} \simeq 32\,\mathrm{km}$ over 1 century. IF conservatively rescaled by a factor of ten, it yields  $\upsigma_r\lesssim 320\,\mathrm{km}$, in substantial agreement with the latest evaluations of the Uranus' orbit uncertainty \citep{2025AJ....169...65J} obtained by reducing astrometry data against the Gaia star cataolog. 

Instead, if the orbit of Uranus were to be determined with the same accuracy as that of Saturn, the situation would change dramatically. Indeed, if one day one could ever count on realistic errors for Uranus' orbital rates of change a hundred times smaller\footnote{It would correspond to a linear uncertainty in the Uranus' orbit of about 32 m over, say, ten years.} than those in Table \ref{Tab:2}, it would be possible to completely exclude the existence of all planetary candidates in the case in which there were still no anomalous orbital signatures statistically different from zero. That is, if even after such a significant improvement in the measurements, the anomalous orbital precessions of Uranus continued to be statistically compatible with zero, repeating the analysis done with Saturn it turns out that there would be no permitted regions at all in the parameter space of the previously examined planetary candidates in the form in which they have been proposed so far. Indeed, allowed regions in the parameter space of, say, a 5 Earth mass body would start to appear only for $a^{'}\gtrsim 750\,\mathrm{au}$, while for $m^{'}=8.4\,m_\oplus$ there would be no room in the perturber's parameter space for $a^{'}\lesssim 950 \,\mathrm{au}$.

Such a prospect of improving our knowledge of Uranus's orbit may not be entirely unrealistic given the great attention that an interplanetary automated mission to the seventh planet of the solar system has been receiving for some years now \citep{2012ExA....33..753A,2014P&SS..104..122A,2014P&SS..104...93T,2015AdSpR..55.2190B,2017AdSpR..59.2407M,2018P&SS..155...12M,cinesi18,2019P&SS..17704680H,2020P&SS..19105030F,
2020Natur.579...17G,2020AcAau.170....6J,2020SSRv..216...72S,2020SSRv..216...22V,
2021ExA...tmp..139G,2021JSpRo..58..505S,2022PSJ.....3...58C,2022Natur.604..607W,2023MNRAS.524L..32B,2023AcAau.202..104G,2023MNRAS.523.3595I,UOP23,2025PhRvD.112h3029Z}.

One might wonder whether the same considerations also apply to Neptune, given that some of the previously cited works for Uranus have also the eight planet of the solar system among its goals along with others dedicated to it and its moon Triton \citep{2021BAAS...53d.371M,2021PSJ.....2..184R}. Actually, it would be difficult to give a positive answer since the orbital period of Neptune is as long as 165 years. This implies that the current record of modern observations, starting in 1913, does not yet cover even a complete orbital revolution of it, with at least another 50 years from now needed. Furthermore, even if any orbiting mission were to actually reach Neptune, for its hopefully accurate observations to have a significant impact on the planet's orbit determination it would have to transmit them continuously for many years. Thus, hoping to be able to use a day Neptune's orbital precessions averaged over its orbital period appears impracticable. The same is even more true for Pluto.
\section{What could be done with a dedicated deep space probe}\label{Sec:6}
\begin{figure}[ht!]
\centering
\begin{tabular}{cc}
\includegraphics[width = 6.5 cm]{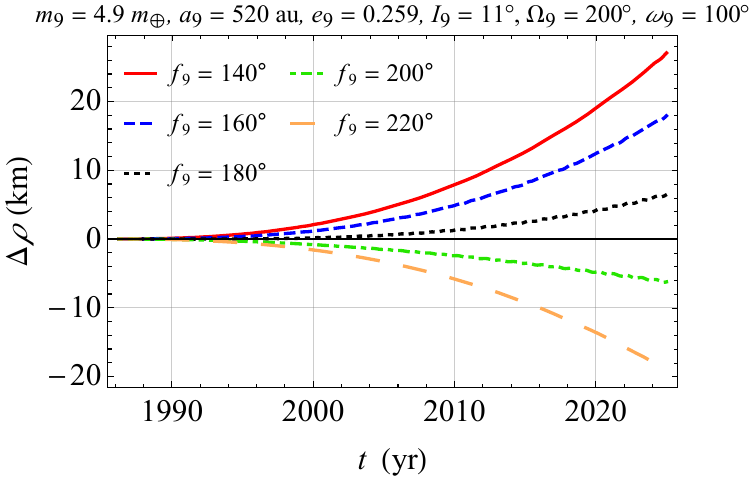} & \includegraphics[width = 6.5 cm]{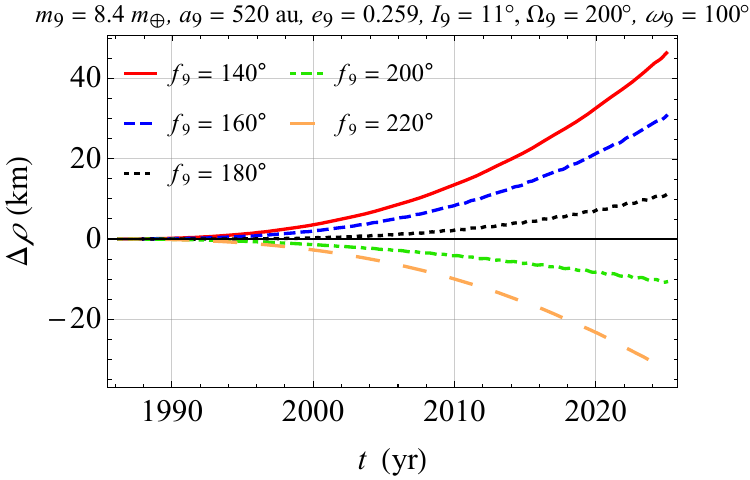}\\
\includegraphics[width = 6.5 cm]{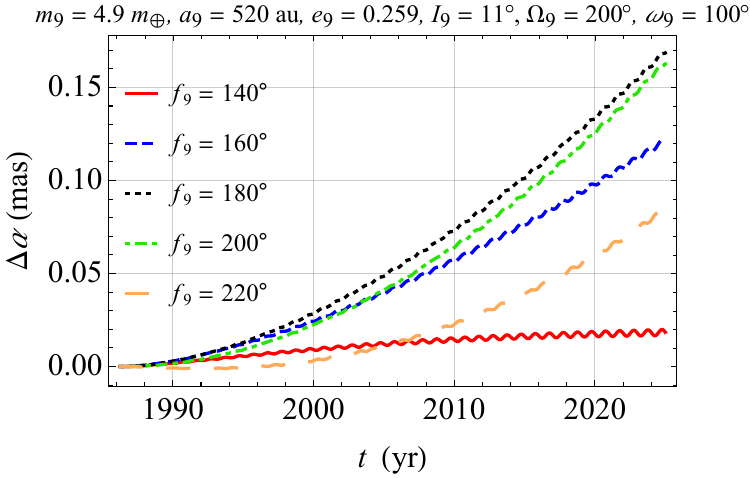} & \includegraphics[width = 6.5 cm]{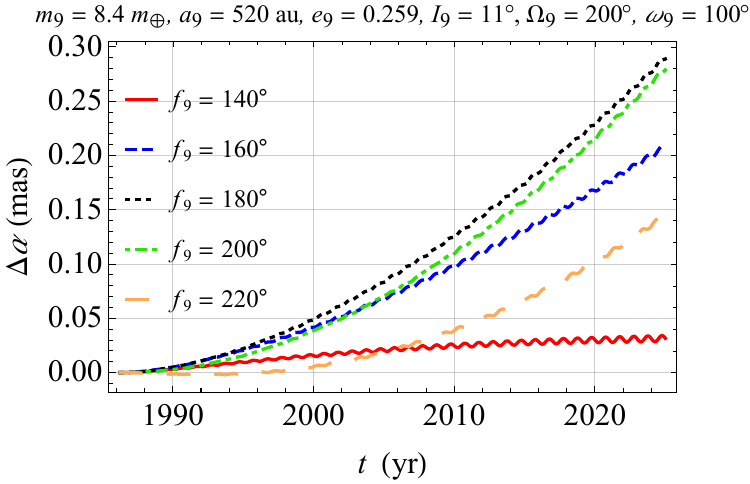}\\
\includegraphics[width = 6.5 cm]{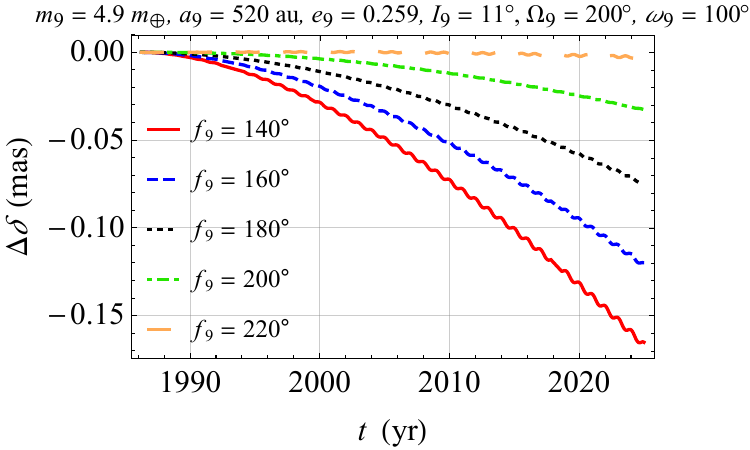} & \includegraphics[width = 6.5 cm]{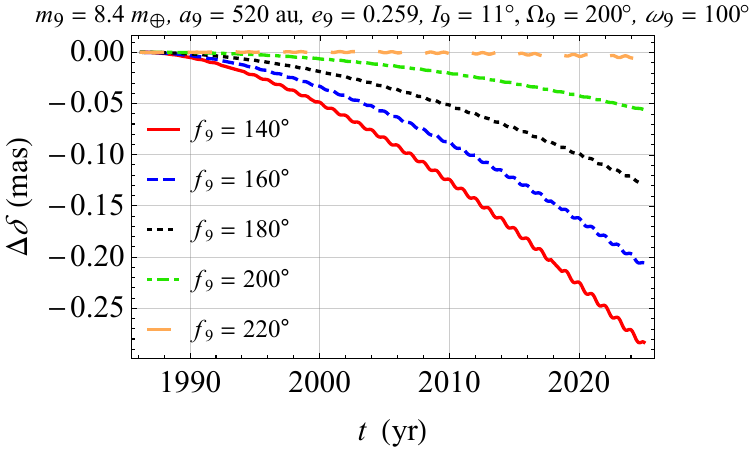}\\
\end{tabular}
\caption{Numerically produced time series $\Delta\rho(t),\Delta\alpha(t),\Delta\delta(t)$ of the signatures induced on the range $\rho$, RA and decl. of Voyager 1 from March 1986 to today by a P9 with $m_9=4.9\,m_\oplus$ (lef column) and $m_9=8.4\,m_\oplus$ (right column). They were obtained by integrating the barycentric equations of motion of the probe and of all the eight known planets of the solar system with respect to the ICRF with and without P9. Both runs shared the gravitational pulls of the  solar system's planets from Mercury to Neptune and the same initial conditions for both the probe and the planets themselves retrieved from the WEB interface of the HORIZONS program maintained by the JPL, NASA.}\label{Fig:5}
\end{figure}

The Voyager 1 probe, launched in 1977, is now at about 170  Astronomical Units (AU),  and it is still transmitting.  It is escaping the solar system at a speed of about
\footnote{See \url{http://voyager.jpl.nasa.gov/mission/fastfacts.html} on the Internet. Consulted 13 December 2025.} 
$3.6$ AU yr$^{-1}$, 35 degrees out of the ecliptic plane to the north. Thus, even just thinking of being able to use its telemetry
to place constraints on the position of the previously examined planetary candidates represents an irresistible and, at least in principle, perfectly well-founded temptation, especially in view of the fact that it seems that its ranging accuracy may have been significantly better than
\footnote{See $\#$3 at \url{http://boards.straightdope.com/sdmb/showthread.php?t=548525} 
on the Internet. Consulted 13 December 2025.} 1 km, being, perhaps, of the order of 10 m in 2010.
Figure \ref{Fig:5} shows the difference of the time series of the range $\rho$, $\alpha$ and $\delta$ of Voyager 1 obtained by numerically integrating its equations of motion with and without P9 over a time span from March 1986, when it was already on its interstellar course after the flyby with Saturn occurred six years before, to today. The minimum and maximum expected values of its mass were chosen for P9, while its orbital parameters approximately correspond to the allowed regions inferred from the Kronian orbital effects in Section \ref{Sec:4}.  
It can be noted that, while the P9-induced astrometric signatures are at the sub-mas level even after 39 years, the range shift, instead, may reach conspicuous values amounting up to about $20-40\,\mathrm{km}$ over the same time span. Incidentally, it would have been as large $\Delta\rho\simeq 5-10\,\mathrm{km}$ in 2010. 
Considering that such large values refer to orbital configurations of P9 permitted by Saturn's orbit, 
they are certainly interesting from the point of view of the potential of Voyager 1 with regard to the constraining of the hypothetical planet.

Unfortunately, such an ideally appealing possibility should be deemed as unfeasible because of a number of reasons exposed below\footnote{W.M. Folkner, private communication, July 2014}. Typically, uncertainties in estimated acceleration of interplanetary spacecraft due to non-gravitational effects such as outgassing of materials or thermal radiation may be quite larger than that due to a distant planetary candidate like, e.g., P9 which is of the order of $\simeq 10^{-13}\,\mathrm{m\,s}^{-2}$ or less. For example, the estimated uncertainty in the acceleration of the Pioneer 10 spacecraft due to thermal emission of its radio isotope thermal electric generator (RTG) is about $2\times 10^{-10}\,\mathrm{m\,s}^{-2}$ \citep{2012PhRvL.108x1101T}. The Pioneer 10 thermal acceleration is especially amenable for estimation because it was a spinning spacecraft far from the Sun. The Voyager spacecraft, instead, are not spin-stabilized,  using thrusters to control attitude which fire in an unbalanced mode, resulting in much larger unknown acceleration  due to thruster usage for attitude control \citep{1983ITAC...28..256C}. The uncertainty in acceleration due to thermal radiation from the RTGs used on Voyager is probably larger than the uncertainty in the effect on Pioneer 10, but in any case much smaller than the acceleration uncertainty due to thrusters.

Nonetheless, Figure \ref{Fig:5} may still be useful if it is thought as representative of the effect of P9 over, say, 39 yr on a hypothetical spin-stabilized deep-space probe, endowed with an accurate ranging apparatus transmitting over a continuous basis over the years, which moves along the same path as Voyager 1 just for illustrative purposes. The idea is not entirely far-fetched, as a similar proposal was put forward a decade ago \citep{2015PhRvD..92j4048B} to test various infrared (IR) modified gravity models, including the Yukawa-type, with a spacecraft at 100 au. In that case, however, the effect of the objects of the Kuiper belt was seen as a major source of disturbance, and to minimize it, a trajectory perpendicular to the ecliptic was chosen.
\section{Summary and conclusions}\label{Sec:7}
The long-term rates of change of the Keplerian orbital elements of a two-body system perturbed by a distant, pointlike massive body were analytically calculated to the Newtonian tidal quadrupolar order in the limiting case of small eccentricity and inclination, an approximation which is valid for solar system's major planets when their orbits are referred to the ecliptic. In order to remove the latter restriction, the calculation was repeated for arbitrary values of the inclination as well. The obtained expressions were averaged over one orbital period of the disturbed planet by assuming that the position in space of the remote perturber does not essentially vary over such a time span.

They were used in conjunction with the current uncertainties in the observationally determined orbital rates of change of Saturn found in the literature to tentatively constrain the parameter space of some hypothetical planetary candidates put forth in recent years to explain certain allegedly observed features of the Kuiper belt. The latter scenarios generally envisage the existence of a still undiscovered planet whose mass is expected to range from approximately one Mercury mass up to 8.4 Earth masses  located at hundreds astronomical units away from the Sun, and whose orbital plane is tilted by about one or two tens of degrees to the ecliptic. In principle, one may not straightforwardly compare analytical formulas of the orbital effects induced by such a putative body with post-fit residual quantities obtained by fitting incomplete dynamical force models, not including also the aforementioned perturber(s), to data records, as, instead, done in this work. This is because there is a possibility that, in the data reduction procedure,  the eventual signal sought, if really existent, will be somehow absorbed in the determined values of the solved-for parameters like, e.g., the planets' state vectors, thus possibly resulting in unrealistically tight constraints. In fact, the planetary ephemerides on which the uncertainties adopted  are based were produced without modeling any of the hypothetical objects considered here. Nonetheless, in order to cope with this potential issue, the formal errors in the Kronian orbital rates of change were rescaled by a factor of ten. Furthermore, the results of this study are not in contrast with other works in which the aforementioned procedure was implemented with real observations.

It turned out that while the latest version of Planet X is provisionally ruled out, there are a few well detached and rather narrow allowed regions only in the parameter space of a Mercury-sized Planet Y for values of its semimajor axis $a_\mathrm{Y}$ of 125 astronomical units; they tend to expand and merge together as the latter increases up to its predicted maximum value of 200 astronomical units. In all such cases, the permitted values of its true anomaly $f_\mathrm{Y}$ tend to lie in ranges broadly corresponding to its aphelion. Instead, a Planet Y of one Earth mass is ruled out. As far as the most recent prediction for Planet Nine is concerned, it is not ruled out. For a body with 4.9 Earth masses, allowed regions occur only for values of its semimajor axis $a_9$ as large as 520 astronomical units. They tend to be more detached and narrow for values of the eccentricity $e_9$ of about 0.2, while they get wider for $e_9\simeq 0.5$. In both cases, the true anomaly $f_9$ is confined between $140^\circ$ and $220^\circ$, thus pointing toward a possible location close to the aphelion. The allowed regions of the parameter space of an object with $8.4$ Earth masses, which exist only for $a_9=520\,\mathrm{au},\,e_9\simeq 0.5$, are well detached and quite narrow, being $f_9$ constrained within about $160^\circ-200^\circ$. In all such cases, the argument of perihelion $\omega_9$ and the longitude of the ascending node $\Omega_9$ of Planet Nine appear constrained within more or less wide and disjointed allowed zones distributed from $0^\circ$ to $360^\circ$. When a different type of the parameter space, based on the ICRF and the usual astrometric angles, is adopted, it turned out that allowed regions exist starting from heliocentric distances $r_9$ as large as 600 and 700 astronomical units for $4.9$ and $8.4$ Earth masses, respectively.

Improving the orbital accuracy of Uranus, currently rather modest because of lacking of long enough records of accurate radiotechnical observations, with a possible interplanetary mission targeted at this gaseous giant would be of great significance for the present scopes. Indeed, if it were possible to push the accuracy of its orbital precessions to the level of Saturn's, thus improving them by about 100 times compared to the present-day level, the allowed regions for a putative body with 5 and 8 Earth masses  would start at no less than 750 and 950 astronomical units, respectively, should no anomalous signatures be recorded.

The trajectory of a hypothetical deep-space spacecraft moving, say, along the same path of Voyager 1 would be perturbed by Planet Nine by some tens of kilometers after about 40 yr, which makes the prospect of implementing this project interesting.

In conclusion, strictly speaking, the allowed zones obtained here may not be viewed as actual constraints on the position of the planetary candidates examined since they were ultimately inferred from planetary ephemerides produced without modeling any additional planet. Rather, they should be viewed more as  hints of what might be detectable should planetary ephemerides actually include them in the dynamical models fitted to observations. Nevertheless, they should be considered credible both because the formal errors used to obtain them were conservatively rescaled by a factor of ten, and because, in the case of Planet Nine, they are substantially in agreement with the constraints obtained independently by other researchers who explicitly modeled it. After all, Le Verrier fitted a purely Newtonian dynamical model to the observations of Mercury, yet its then anomalous perihelion precession remained intact without the general relativistic signal being absorbed into the estimated parameters.
\section*{Data availability}
No new data were generated or analysed in support of this research.
\section*{Conflict of interest statement}
I declare no conflicts of interest.
\section*{Funding}
This research received no external funding.
\section*{Acknowledgements}
I am grateful to Amir Siraj and William M. Folkner for important discussions and information which notably contributed to improve this paper.

\bibliography{Megabib}{}
\end{document}